\documentclass[preprint,amsmath,amssymb,nofootinbib,floatfix]{revtex4-2}

\usepackage{graphicx}
\usepackage{dcolumn}
\usepackage{bm}
\usepackage[utf8]{inputenc}
\usepackage[T1]{fontenc}
\usepackage{mathptmx}
\usepackage{booktabs}
\usepackage{xcolor}
\usepackage{placeins}
\usepackage{hyperref}
\hypersetup{
    unicode,
    colorlinks=true,
    linkcolor=blue,
    citecolor=blue,
    urlcolor=blue
}

\begin{document}

\title{Multi-diagnostic convergence: a single measurement in weakly collisional plasmas}

\author{Victor Edmonds}
\email{vedmonds@finalstopconsulting.com}
\affiliation{Final Stop Consulting LLC, Fuquay-Varina, NC, United States}

\begin{abstract}
When multiple electron temperature diagnostics converge on the same value, the standard inference is that the measurement is robust. We show that this convergence is a structural consequence of the shared ionization bottleneck in any plasma where the electron Knudsen number exceeds $\sim\!0.01$: all diagnostics downstream of collisional ionization report the effective temperature $T_{\rm eff}$ of the electron velocity distribution, not the core temperature $T_{\rm core}$. Their agreement is a single measurement reported $N$ times. We introduce a diagnostic taxonomy classifying methods as ionization-gated (Type~A, measures $T_{\rm eff}$), bulk-sampling (Type~B, measures $T_{\rm core}$), or distribution-resolving (Type~C). The ratio $R = T_A/T_B$ yields $\kappa = 3R/[2(R-1)]$ directly. We apply the framework to the solar corona ($R = 2.4$, $\kappa \approx 2.5$) and validate it quantitatively in the tokamak scrape-off layer, where single kappa distributions ($\kappa \approx 2$--10) reproduce published bi-Maxwellian EEDF decompositions to 3--8\% RMS with one fewer free parameter and Thomson scattering independently confirms the predicted Type~B temperature. We test the framework's boundary of applicability in planetary nebulae (the 80-year CEL--ORL abundance discrepancy). Knudsen number calculations, including the Shoub $v^4$ mean-free-path scaling for tail electrons, show that the ionizing population is collisionless in the corona even when the bulk is fluid; in planetary nebulae, both the ionizing electrons ($\sim$55~eV) and the lower-energy excitation electrons ($\sim$5~eV) that drive the CEL diagnostic are collisional over nebular scales, identifying PNe as the falsification boundary of the framework; in the tokamak SOL, non-local parallel transport maintains suprathermal tails even where local collisionality is high. For plasmas with $\kappa \approx 3$--5, the raw Spitzer--H\"arm formula with spectroscopic $T_e$ as input overestimates parallel heat flux by factors of 3--25$\times$; flux-limited transport models inherit the temperature bias through their boundary conditions, with direct relevance to ITER divertor predictions. Every diagnostic campaign on a weakly collisional plasma should include at least one Type~B measurement.
\end{abstract}

\maketitle

\section{Introduction}
\label{sec:intro}

When five independent electron temperature diagnostics converge on the same value, the standard inference is that the measurement is robust. In the quiet solar corona, five EUV-based methods agree on $T_e \approx 1.5$~MK \cite{DelZanna2018}. In the tokamak scrape-off layer, spectroscopic line ratios are routinely cross-validated against emission measure analysis. In planetary nebulae, abundances derived from collisionally excited lines and recombination lines have been compared for over 80 years \cite{Wyse1942}, with persistent discrepancies later formalized as the abundance discrepancy factor \cite{Peimbert1967}. The assumption underlying all of these: the electron velocity distribution is Maxwellian.

This paper demonstrates that for any plasma where the electron Knudsen number exceeds $\sim$0.01 \cite{ScudderKarimabadi2013}, the convergence is a structural consequence of the shared ionization bottleneck, regardless of the distribution shape. All ionization-based diagnostics are downstream of a single bottleneck: collisional ionization, a threshold-crossing process dominated by the suprathermal tail \cite{Owocki1983}. For non-Maxwellian distributions, the ionization rate tracks the effective temperature $T_{\rm eff}$ rather than the core temperature $T_{\rm core}$, and every diagnostic that depends on charge state populations inherits this bias. The agreement measures $T_{\rm eff}$. It cannot detect a departure from Maxwellian because the departure is invisible to the measurement channel.

The result is general. It depends on the mathematics of threshold-crossing integrals and applies across plasma environments and generation mechanisms. We introduce a diagnostic taxonomy (Type~A: ionization-gated; Type~B: bulk-sampling; Type~C: distribution-resolving) that classifies which measurements are degenerate in the distribution shape parameter $\kappa$ and which break the degeneracy. The ratio of any Type~A to any Type~B temperature applied to the same plasma yields $\kappa$ directly: $\kappa = 3R/[2(R-1)]$, where $R = T_A/T_B$.

The motivating example is the solar corona, where the diagnostic temperature discrepancy ($R = 2.4 \pm 0.3$) implies $\kappa \approx 2$--3 \cite{Mercier2015}. The present paper generalizes this result, demonstrates its applicability to the tokamak scrape-off layer and to planetary nebulae, and quantifies the consequences for heat transport predictions.

This paper does not claim that all reported plasma temperatures are wrong. In high-density, collision-dominated plasmas ($\mathrm{Kn} \ll 0.01$), the Maxwellian assumption is recovered and ionization-based diagnostics are reliable. The framework identifies where the assumption fails and by how much.

\subsection{Plan of the paper}

Section~\ref{sec:framework} develops the mathematical framework: the kappa distribution, the tail-dominance of ionization, the insensitivity of excitation, the circular convergence principle, and its generalization to bi-Maxwellian distributions. Section~\ref{sec:taxonomy} introduces the diagnostic taxonomy. Section~\ref{sec:domains} applies the framework to three domains: the solar corona (Sec.~\ref{sec:corona}), the tokamak scrape-off layer (Sec.~\ref{sec:sol}), and planetary nebulae (Sec.~\ref{sec:nebulae}), with additional environments summarized in Sec.~\ref{sec:other_domains}. Section~\ref{sec:break} discusses which diagnostics break the degeneracy. Section~\ref{sec:implications} quantifies implications for heat transport and presents the diagnostic prescription. Section~\ref{sec:discussion} addresses anticipated criticisms, relation to existing work, and open questions.

\section{Mathematical Framework}
\label{sec:framework}

\subsection{Kappa distributions}
\label{sec:kappa_recap}

Non-equilibrium plasmas are characterized by kappa distributions \cite{Vasyliunas1968,Tsallis1988,Livadiotis2009,Livadiotis2011,Pierrard2010}, which exhibit power-law tails rather than the exponential cutoff of a Maxwellian:
\begin{equation}
f_\kappa(v) \propto \left(1 + \frac{v^2}{\kappa \theta^2}\right)^{-(\kappa+1)}
\label{eq:kappa}
\end{equation}
where $\theta^2 = 2k_BT_{\rm core}/m$ and $\kappa$ parameterizes the departure from equilibrium. As $\kappa \to \infty$, the Maxwellian is recovered. At low $\kappa$, the distribution has an enhanced high-energy tail.

The effective temperature, defined via the second moment, is:
\begin{equation}
T_{\rm eff} = T_{\rm core} \cdot \frac{\kappa}{\kappa - 3/2}
\label{eq:amplification}
\end{equation}
$T_{\rm eff}$ governs processes sensitive to the mean kinetic energy (pressure, scale height) or to the suprathermal population (ionization, recombination). $T_{\rm core}$ governs processes dominated by the thermal peak.\footnote{We use the ``Kappa-A'' formulation of \protect\citet{Lazar2016} in which $T_{\rm core}$ is the most-probable-speed temperature. The ``Kappa-B'' formulation defines $T = T_{\rm eff}$ throughout. All physical results are independent of the convention chosen; see \protect\citet{Lazar2016} for a comprehensive comparison. For $\kappa \leq 5/2$, the standard kappa distribution has divergent higher moments; the regularized formulation of \protect\citet{Scherer2017} resolves this while preserving the power-law tail at moderate energies.}

\subsection{Why ionization is tail-dominated}
\label{sec:tail_dominance}

Collisional ionization has threshold energies $E_{\rm th}$ that typically far exceed $k_BT_{\rm core}$. For diagnostic ions in the three environments considered in this paper:
\begin{itemize}
\item Solar corona: Fe~IX--XII ionization potentials 234--331~eV at $T_{\rm core} \approx 52$~eV ($E_{\rm th}/k_BT_{\rm core} \approx 4.5$--6.4)
\item Tokamak SOL: C~III at 47.9~eV, $T_{\rm core} \sim 5$--30~eV ($E_{\rm th}/k_BT_{\rm core} \approx 1.6$--10)
\item Planetary nebulae: O~III at 54.9~eV, $T_{\rm core} \approx 0.86$~eV ($E_{\rm th}/k_BT_{\rm core} \approx 64$)
\end{itemize}

Ionization cross-sections rise above threshold, so the rate coefficient integrand peaks in the tail. \citet{Owocki1983} demonstrated that for non-Maxwellian distributions, ionization rates of high-threshold ions are controlled by the suprathermal population rather than the thermal core; the effect diminishes for low-threshold species where the ionization cross-section samples nearer the thermal peak. Comprehensive ionization equilibrium calculations for kappa distributions across H through Zn were provided by \citet{Dzifcakova2013}, with synthetic spectral predictions enabled by the KAPPA software package \cite{Dzifcakova2015KAPPA}.

\subsubsection{Local slope temperature}

The logarithmic slope of the kappa energy distribution at energy $E$ is
\begin{equation}
\frac{d \ln f_\kappa}{dE} = \frac{1}{2E} - \frac{\kappa + 1}{\kappa\, k_B T_{\rm core} + E}\,.
\label{eq:log_slope}
\end{equation}
For a Maxwellian at temperature $T$, the corresponding slope is $1/(2E) - 1/(k_BT)$. Equating the distribution-dependent terms defines the \emph{local slope temperature}, the Maxwellian temperature that matches the kappa distribution's gradient at energy $E$:
\begin{equation}
T_{\rm loc}(E) = \frac{\kappa\, k_B T_{\rm core} + E}{(\kappa + 1)\, k_B}\,.
\label{eq:Tloc}
\end{equation}
This quantity has three important properties. First, $T_{\rm loc}(0) = \kappa T_{\rm core}/(\kappa+1) < T_{\rm core}$: the core of a kappa distribution is steeper than any Maxwellian at $T_{\rm core}$, containing fewer moderate-energy electrons. Second, $T_{\rm loc}$ increases linearly with $E$: the kappa tail becomes progressively shallower at higher energies, mimicking a hotter Maxwellian the further from the thermal peak one samples. Third, $T_{\rm loc} \to T_{\rm core}$ as $\kappa \to \infty$ at every energy, recovering the Maxwellian limit.

\subsubsection{Rate equivalence}

\begin{figure}[!htbp]
\centering
\includegraphics[width=\columnwidth]{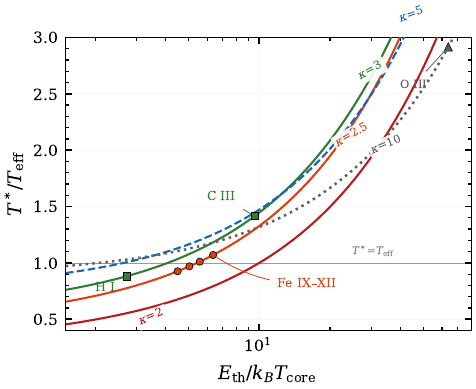}
\caption{Rate-equivalent Maxwellian temperature $T^*$ normalized to $T_{\rm eff}$, as a function of the ionization threshold $E_{\rm th}/k_BT_{\rm core}$, for kappa distributions with $\kappa = 2$--10. Filled circles mark diagnostic ions in each domain: Fe~IX--XII for the solar corona ($\kappa = 2.5$; ionization thresholds from NIST \cite{NIST_ASD}), H~I and C~III for the tokamak SOL ($\kappa = 3$), and O~III for planetary nebulae ($\kappa = 10$). Coronal parameters ($n_e = 10^8$~cm$^{-3}$, $T_{\rm core} = 52$~eV) follow \citet{Mercier2015}; SOL parameters ($T_{\rm core} = 5$~eV) are representative of far-SOL conditions \cite{Stangeby2000}; PN parameters ($T_{\rm core} = 0.86$~eV) follow standard nebular values \cite{Wesson2018}. The coronal diagnostic ions cluster tightly around $T^*/T_{\rm eff} \approx 1$ (dashed line), confirming that $T_{\rm eff}$ is an excellent proxy for the rate-equivalent temperature in that regime. At high thresholds (PN regime), $T^*$ substantially exceeds $T_{\rm eff}$, and the inversion formula [Eq.~(\ref{eq:inversion})] becomes quantitatively unreliable.}
\label{fig:rate_equivalence}
\end{figure}

The ionization rate coefficient for electron impact from the ground state of ion $X^{q+}$ is
\begin{equation}
\langle \sigma_q v \rangle = \int_{E_{\rm th}}^{\infty} \sigma_q(E)\, v(E)\, F(E)\, dE\,,
\end{equation}
where $F(E)$ is the electron energy distribution and $\sigma_q(E)$ vanishes below the threshold $E_{\rm th}$. Because $\sigma_q$ turns on at $E_{\rm th}$ and $F$ decays above it, the integrand is dominated by energies near $E_{\rm th}$. Following the saddle-point framework of \citet{Owocki1983}, the integral is controlled by the local slope of $F$ at $E_{\rm th}$. To the extent that the kappa distribution near $E_{\rm th}$ resembles a Maxwellian at $T_{\rm loc}(E_{\rm th})$, the rate coefficient satisfies
\begin{equation}
\langle \sigma_q v \rangle_\kappa(T_{\rm core}) \approx \langle \sigma_q v \rangle_{\rm Max}\!\bigl(T_{\rm loc}(E_{\rm th})\bigr)\,.
\label{eq:rate_equivalence}
\end{equation}
Numerical integration using Lotz-type cross-sections\footnote{The \citet{Lotz1967} empirical formula provides an adequate approximation for the rate-equivalence argument; more accurate distorted-wave or $R$-matrix cross-sections would shift~$T^*$ by $\lesssim$10\% without affecting the convergence conclusion.} confirms that $T_{\rm loc}$ predicts the rate-equivalent Maxwellian temperature to within ${\sim}\,10$--$30\%$ across the parameter space relevant to each domain (Table~\ref{tab:rate_check}; Fig.~\ref{fig:rate_equivalence}).

\begin{table}[htbp]
\centering
\caption{Rate-equivalent Maxwellian temperature $T^*$ (from numerical integration with Lotz cross-sections) compared to the local slope prediction $T_{\rm loc}$ and to $T_{\rm eff}$. The ratio $T^*/T_{\rm eff}$ shows that the $T_{\rm eff}$ approximation is excellent for the solar corona and moderate elsewhere. The final column gives $\kappa_{\rm inferred}$ obtained by applying the inversion formula [Eq.~(\ref{eq:inversion})] to $R = T^*/T_{\rm core}$, with the percentage error relative to the true $\kappa$.}
\label{tab:rate_check}
\begin{ruledtabular}
\begin{tabular}{llcccccc}
Domain & Ion & $E_{\rm th}/k_BT_{\rm core}$ & $\kappa$ & $T^*/T_{\rm core}$ & $T^*/T_{\rm eff}$ & $T^*/T_{\rm loc}$ & $\kappa_{\rm inf}$ (error) \\
\hline
Corona  & Fe~IX  & 4.5 & 2.5 & 2.31 & 0.93 & 1.16 & 2.64 (+5.6\%) \\
Corona  & Fe~XII & 6.4 & 2.5 & 2.68 & 1.07 & 1.06 & 2.39 ($-$4.3\%) \\
SOL     & H~I    & 2.7 & 3.0 & 1.76 & 0.88 & 1.23 & 3.47 (+16\%) \\
SOL     & C~III  & 9.6 & 3.0 & 2.83 & 1.42 & 0.89 & 2.32 ($-$23\%) \\
PN      & O~I    & 15.8 & 10 & 1.79 & 1.52 & 0.76 & 3.40 ($-$66\%) \\
PN      & O~III  & 63.8 & 10 & 3.43 & 2.92$^{\rm a}$ & 0.51 & 2.12$^{\rm a}$ ($-$79\%) \\
\end{tabular}
\end{ruledtabular}
\smallskip
\noindent$^{\rm a}$The inversion formula is inapplicable at this extreme threshold ratio; see Sec.~\ref{sec:nebulae} for why the [O~III] diagnostic requires forward modeling rather than the simple inversion.
\end{table}

\begin{figure}[htbp]
\centering
\includegraphics[width=\columnwidth]{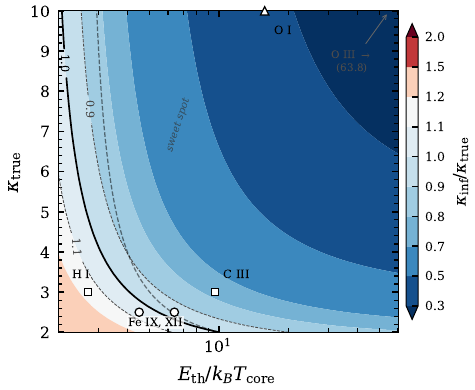}
\caption{Systematic error of the inversion formula: $\kappa_{\rm inferred}/\kappa_{\rm true}$ across the $(E_{\rm th}/k_BT_{\rm core},\, \kappa)$ parameter space, computed by numerical integration of kappa rate coefficients with Lotz cross-sections. The ``sweet spot'' curve [Eq.~(\ref{eq:sweet_spot})] marks where $T^* = T_{\rm eff}$ and the inversion is exact. Filled markers show diagnostic ions from Table~\ref{tab:rate_check}. The coronal Fe~IX--XII ions cluster near the sweet spot ($<$10\% error); SOL ions (H~I, C~III) show 15--25\% error; PN ions are off-scale. The diagnostic prescription (\S\ref{sec:prescription}) references this figure for the validity domain.}
\label{fig:error_propagation}
\end{figure}

The systematic error of the inversion formula is mapped across the full parameter space in Fig.~\ref{fig:error_propagation}. The formula recovers $\kappa$ to better than 10\% in the coronal sweet-spot band (Fe~IX--XII), degrades to 15--25\% in the SOL regime (H~I, C~III), and fails by factors of 2--5 at extreme threshold ratios (PN regime). The observational uncertainty propagation is $\delta\kappa = 3\,\delta R/[2(R-1)^2]$; for the coronal measurement $R = 2.4 \pm 0.3$, this gives $\delta\kappa \approx \pm 0.23$ ($\pm 9\%$), comparable to the systematic approximation error. For the tokamak SOL at $R \approx 1.45 \pm 0.3$, observational uncertainty dominates ($\delta\kappa \approx \pm 2.2$, or $\pm 46\%$): the inversion formula is a coarse diagnostic in the SOL, not a precision measurement. Both systematic and statistical errors are now explicit in Table~\ref{tab:rate_check}.

\FloatBarrier

A special case: $T_{\rm loc}(E_{\rm th}) = T_{\rm eff}$ when
\begin{equation}
\frac{E_{\rm th}}{k_BT_{\rm core}} = \frac{5\kappa}{2\kappa - 3}\,.
\label{eq:sweet_spot}
\end{equation}
For $\kappa = 2.5$, this gives $E_{\rm th}/k_BT_{\rm core} = 6.25$; for $\kappa = 3$, it gives 5.0. The solar coronal diagnostic ions Fe~IX--XII have $E_{\rm th}/k_BT_{\rm core} \approx 4.5$--$6.4$, straddling this value. Their mean rate-equivalent temperature is $T^* = 0.994\,T_{\rm eff}$, confirming that $T_{\rm eff}$ is an excellent proxy in the solar case.

More generally, the rate-equivalent temperature $T^*$ exceeds $T_{\rm eff}$ at high thresholds (where the power-law tail dominates) and falls below it at low thresholds (where the distribution still resembles the core). The saddle-point approximation underlying Eq.~(\ref{eq:rate_equivalence}) degrades at extreme threshold ratios ($E_{\rm th}/k_BT_{\rm core} \gg 10$) because the kappa power-law tail becomes so flat relative to the exponential assumed in the saddle-point expansion that a single local slope no longer captures the integral; the rate is then controlled by a broad range of energies above $E_{\rm th}$, not a narrow saddle near it. Because the inversion formula [Eq.~(\ref{eq:inversion})] is decreasing in $R$, the overestimation of $T^*$ at high threshold ratios propagates to a systematic undershoot of $\kappa_{\rm inferred}$, visible for C~III in the SOL and O~I/O~III in PNe (Table~\ref{tab:rate_check}). The \citet{Owocki1983} principle (ionization tracks a temperature above $T_{\rm core}$) is robust; the specific value depends on $E_{\rm th}/k_BT_{\rm core}$.

\subsubsection{Convergence across ions}

For the circular convergence argument, the critical question is not whether $T^* = T_{\rm eff}$ exactly, but whether multiple diagnostics converge on the \emph{same} $T^*$. Since ionization equilibrium is set by the balance of all ionization and recombination rate coefficients (each an integral over the same distribution $f_\kappa$), the resulting charge state fractions are a single self-consistent set. Any diagnostic downstream of these fractions maps them to the same Maxwellian-equivalent temperature. The inter-ion spread in $T^*$ quantifies how tight the convergence is.

For the solar corona ($\kappa = 2.5$), Fe~IX through Fe~XII span $T^* = (2.31$--$2.68)\,T_{\rm core}$, a range of 14.5\% around the mean, comparable to the $\sim$20\% absolute calibration uncertainty of current EUV instruments. The diagnostics appear to converge. They are measuring a single quantity: $T_{\rm eff}$ (to good approximation), not $T_{\rm core}$.

This equivalence is not specific to kappa distributions. Any distribution with an enhanced suprathermal tail will produce ionization rates exceeding the Maxwellian prediction at $T_{\rm core}$. Kappa is the analytically tractable case; the convergence principle is distribution-agnostic. \citet{HahnSavin2015} demonstrated independently that kappa ionization rate coefficients can be reproduced to $<$3\% accuracy by a weighted sum of Maxwellians, confirming that the mapping from kappa to a Maxwellian-equivalent temperature is robust and not an artifact of any particular integration technique.

\subsection{Why excitation is not tail-dominated}
\label{sec:excitation}

Excitation energies for allowed transitions within a given ion are comparable to $k_BT$: typical strong coronal EUV lines have excitation thresholds of $E_{\rm exc} \sim 2$--$5\,k_BT$, placing the rate-determining electrons near the thermal peak. Ionization thresholds, by contrast, sit at $E_{\rm ion} \sim 15$--$40\,k_BT$ for the same ions, deep in the suprathermal tail. This order-of-magnitude separation is why excitation rates are nearly distribution-insensitive while ionization rates are not. \citet{Dudik2014} computed synthetic spectra for kappa distributions and confirmed that the vast majority of strong coronal EUV line ratios vary by less than 20\% across $\kappa = 2$--25.

Once ionization sets the charge states at $T_{\rm eff}$, line ratios within each ion cannot detect the departure. The emission measure of a given ion is proportional to its ion fraction (set by $T_{\rm eff}$) multiplied by excitation rates (nearly $\kappa$-independent). Every ionization-gated diagnostic therefore converges on $T_{\rm eff}$, regardless of whether the measurement exploits line ratios, absolute intensities, or differential emission measure reconstruction.

This argument applies when $E_{\rm exc} \lesssim 5\,k_BT_{\rm core}$, which holds for diagnostic ions in hot plasmas (coronal EUV lines, tokamak upstream). In cold plasmas where $E_{\rm exc} \gg k_BT_{\rm core}$, excitation itself becomes a threshold-crossing process dominated by the tail. Helium emission spectroscopy in the tokamak divertor ($E_{\rm exc} \approx 21$--23~eV at $T_e \sim 2$~eV, giving $E_{\rm exc}/k_BT \approx 10$) and the [O~III] $\lambda 4363$ diagnostic in planetary nebulae ($E_{\rm exc} \approx 5.35$~eV at $T_{\rm core} \approx 0.86$~eV, $E_{\rm exc}/k_BT \approx 6$) both fall in this regime. Same-ion line ratios in these environments track $T_{\rm loc}(E_{\rm exc})$ rather than $T_{\rm core}$ or $T_{\rm eff}$.

Exceptions in the coronal regime also exist. \citet{Dudik2014} found that certain forbidden lines (notably the Fe~X 6378~\AA\ red coronal line, whose intensity increases with $\kappa$) show sensitivity up to a factor of two in temperature-sensitive Fe~IX line ratios across $\kappa = 2$--25. Such transitions, with unusual temperature sensitivity arising from excitation cross-sections that weight the tail more heavily than typical allowed lines, are the spectroscopic channels that could break the degeneracy without requiring a Type~B measurement.

\subsection{The circular convergence principle}
\label{sec:theorem}

\medskip
\noindent\textbf{Principle (Circular convergence).} \textit{Let $f(v)$ be any isotropic electron velocity distribution in a plasma where the charge state distribution is governed by collisional ionization and recombination. Let $\{D_1, D_2, \ldots, D_N\}$ be $N$ diagnostic methods each depending on the electron temperature only through the charge state fractions. If intra-ion excitation rates are approximately independent of the distribution shape (Sec.~\ref{sec:excitation}), then all $N$ diagnostics converge on a narrow band of Maxwellian-equivalent temperatures $T^*$ near $T_{\rm eff}$, regardless of the true form of $f(v)$.}

\medskip
\noindent\textit{Proof sketch.}\quad The ionization equilibrium is uniquely determined by the ionization and recombination rate coefficients $S_q[f]$ and $\alpha_q[f]$, which are functionals of~$f(v)$. Given~$f$, the equilibrium ion fractions are fixed. Any diagnostic downstream of these fractions, when interpreted under Maxwellian tables, maps them to the same~$T^*$. The convergence is exact in the limit that intra-ion excitation rates are $\kappa$-independent; the residual inter-ion spread (Table~\ref{tab:rate_check}) quantifies the approximation.~$\square$

\medskip
Diagnostics conventionally treated as ``independent'' are downstream of a single bottleneck. The convergence is interpreted as cross-validation but does not constitute it.

The logical content of this principle is implicit in the analysis of \citet{Owocki1983}, who demonstrated that ionization rates in kappa plasmas track $T_{\rm eff}$ rather than $T_{\rm core}$; the convergence of ionization-gated diagnostics follows directly. The contribution of the present work is not the logical observation itself but its consequences: the diagnostic taxonomy, the inversion formula, and the cross-domain demonstration that the same bottleneck operates across solar, fusion, and astrophysical plasmas.

\medskip
\noindent\textbf{Corollary 1.} The convergence of $N$ ionization-gated diagnostics contains exactly one independent temperature measurement. Adding method $D_{N+1}$ that is also ionization-gated provides no new information about the distribution shape.

\medskip
\noindent\textbf{Corollary 2.} The only diagnostics capable of detecting the departure are those whose sensitivity function peaks at $v \sim v_{\rm thermal}$ (Type~B) or those that resolve the distribution shape directly (Type~C).

\medskip
\noindent\textbf{Corollary 3 (Inversion formula).} For a kappa distribution where $T^* \approx T_{\rm eff}$ (Sec.~\ref{sec:tail_dominance}), the ratio $R = T_A / T_B$ (where $T_A$ is any Type~A and $T_B$ any Type~B diagnostic applied to the same plasma) yields
\begin{equation}
\kappa = \frac{3R}{2(R-1)}\,.
\label{eq:inversion}
\end{equation}
This approximation is best when the diagnostic ions have thresholds near the value given by Eq.~(\ref{eq:sweet_spot}); for the solar coronal Fe~IX--XII lines, the approximation error is $<$1\% (Table~\ref{tab:rate_check}). For ions with thresholds far from this value, the inversion formula provides an order-of-magnitude estimate of $\kappa$.

\subsection{Distribution generality: bi-Maxwellian example}
\label{sec:bimax}

The convergence principle (Sec.~\ref{sec:theorem}) is stated for arbitrary isotropic $f(v)$ and does not depend on the kappa parametrization. To demonstrate this explicitly, we derive the convergence for bi-Maxwellian distributions: the functional form the fusion community already uses to describe measured EEDFs in the tokamak SOL.

For $f_{\rm biMax}(v) = (1 - f_h)\, F_{\rm Max}(v;\, T_c) + f_h\, F_{\rm Max}(v;\, T_h)$ with $T_h \gg T_c$ and $f_h \ll 1$, the ionization rate coefficient decomposes exactly by linearity:
\begin{equation}
\langle \sigma v \rangle_{\rm biMax} = (1 - f_h)\,\langle \sigma v \rangle_{\rm Max}(T_c) + f_h\,\langle \sigma v \rangle_{\rm Max}(T_h)\,.
\label{eq:bimax_rate}
\end{equation}
For any diagnostic ion with $E_{\rm th} \gg k_BT_c$, the cold component is exponentially suppressed ($\langle \sigma v \rangle_{\rm Max}(T_c) \propto e^{-E_{\rm th}/k_BT_c} \approx 0$) and the rate is dominated by the hot component. Every ionization-gated diagnostic therefore reports a temperature controlled by $T_h$, producing the same circular convergence as in the kappa case.

Thomson scattering fits a Gaussian to the scattered spectrum. For a bi-Maxwellian, the spectrum is a sum of two Gaussians with widths proportional to $\sqrt{T_c}$ and $\sqrt{T_h}$. The narrow cold component dominates the peak shape; the hot component contributes a broad, low-amplitude pedestal. A single-Gaussian fit reports $T_{\rm fit} \approx T_c$, with $T_h$ contributing a small upward bias. At COMPASS parameters ($f_h \approx 0.10$, $T_h/T_c \approx 3$; \citeauthor{Popov2015} 2015), the hot component contributes 6\% of the Thomson spectral amplitude at line center, shifting the single-Maxwellian fit by $+8\%$ above $T_c$---small compared to the $R - 1 \approx 0.45$ signal from the kappa-driven diagnostic bias (cf.\ Sec.~\ref{sec:sol}).

The bi-Maxwellian case is the natural choice for the fusion community: it is the distribution already measured in the CASTOR and COMPASS EEDF decompositions (Sec.~\ref{sec:kappa_fit}), and the rate decomposition [Eq.~(\ref{eq:bimax_rate})] is exact (no saddle-point approximation needed). The kappa framework provides a convenient single-parameter characterization of the resulting diagnostic bias; the convergence principle does not depend on it.

\section{Diagnostic Taxonomy}
\label{sec:taxonomy}

\begin{table*}[t]
\centering
\caption{Diagnostic taxonomy. Type~A methods are ionization-gated and report $T_{\rm eff}$. Type~B methods sample the distribution core and report $T_{\rm core}$ or an intermediate value. Type~C methods resolve the distribution shape directly. A* denotes methods with weak but nonzero $\kappa$ sensitivity, insufficient to break the degeneracy in practice. $^\dagger$Same-ion line ratios cancel the ion fraction and depend purely on excitation rate coefficients. In hot plasmas ($E_{\rm exc} \lesssim 5\,k_BT$), excitation is nearly $\kappa$-independent and the diagnostic converges on $T_{\rm eff}$ through the emission measure. In cold plasmas ($E_{\rm exc} \gg k_BT$), excitation is tail-dominated and the ratio tracks $T_{\rm loc}(E_{\rm exc})$; see Sec.~\ref{sec:excitation}.}
\label{tab:taxonomy}
\begin{ruledtabular}
\begin{tabular}{lclll}
Method & Type & What it measures & $\kappa$ sens. & Domains \\
\hline
Spectral line ratios (same ion)   & A$^\dagger$  & Regime-dependent$^\dagger$        & Regime-dep.$^\dagger$    & Solar, tok., neb. \\
Emission measure / DEM            & A  & $T_{\rm eff}$ (via charge states)        & Blind    & Solar, neb. \\
Ionization equilibrium models     & A  & $T_{\rm eff}$ (by definition)            & Blind    & Solar, neb. \\
CXRS                              & A  & $T_{\rm eff}$ (via ion populations)      & Blind    & Tokamak \\
CELs                              & A* & $T_{\rm eff}$ or $T_{\rm loc}$ (Sec.~\ref{sec:nebulae}) & Weak    & Nebulae \\
Density-sensitive line ratios     & A* & Weakly $\kappa$-dependent                & Mostly blind & Solar, neb. \\
Forbidden line ratios (specific)  & A* & Weakly $\kappa$-sensitive                & Partial  & Solar, neb. \\
\hline
Thomson scattering \cite{Sheffield2011} & B  & $T_{\rm core}$ (distribution width)      & Sensitive & Tok., laser \\
Radio bremsstrahlung (opt.\ thick) & B  & $T_{\rm core}$                           & Sensitive & Solar \\
ORLs                              & B  & $\sim T_{\rm core}$ (recomb.\ $\propto E^{-1}$) & Sensitive & Nebulae \\
Langmuir probes (I-V)             & A/C$^{\ddagger}$ & Sheath-gated or full $f(v)$             & Method-dep. & Tok., lab \\
\hline
In situ particle detectors        & C  & Full $f(v)$                              & Direct   & Solar wind, mag. \\
\end{tabular}
\end{ruledtabular}
\end{table*}

At high densities where $\mathrm{Kn} \to 0$, Coulomb collisions enforce a Maxwellian, all diagnostics recover mutual agreement, and the taxonomy becomes irrelevant; this is the framework's built-in self-consistency check.

Optical recombination lines (ORLs) are Type~B because radiative recombination (RR) cross-sections scale as approximately $\sigma \propto E^{-1}$ above threshold, so the rate coefficient is dominated by slow electrons near the thermal velocity, not the suprathermal tail. Recombination preferentially samples the distribution core, making ORLs the inverse of ionization-gated diagnostics. ORL-derived temperatures therefore approximate $T_{\rm core}$ for kappa distributions, while CEL-derived temperatures track $T_{\rm eff}$ (in CIE environments) or $T_{\rm loc}(E_{\rm exc})$ (in PIE environments where the excitation threshold exceeds $\sim 5\,k_BT$; see Sec.~\ref{sec:nebulae}).

A potential complication is dielectronic recombination (DR), which is a resonant threshold-crossing process that can dominate the total recombination rate for highly charged coronal ions. Because DR requires electrons at specific kinetic energies, it is in principle tail-sensitive. However, for the low-charge ions relevant to the nebular abundance discrepancy (O$^{2+}$, C$^{2+}$, N$^{2+}$) at nebular temperatures ($\sim$10$^4$~K), RR dominates over DR, and the Type~B classification holds. In the solar corona, DR dominates the total recombination rate for many ions and therefore directly affects the ionization equilibrium itself: the charge state balance $N_{q+1}/N_q = S_{\rm ion}/\alpha_{\rm DR}$ has tail-sensitive rates on both sides. Full kappa ionization equilibrium calculations \cite{Dzifcakova2013} show that the ionization rate enhancement outpaces the DR enhancement, so the net shift toward higher charge states is preserved, but it is partially moderated. The convergence principle still holds because the ion fractions remain set by a single tail-dominated bottleneck; the quantitative value of $T^*$ absorbs the DR correction. ORLs are not used as coronal diagnostics, so the Type~B classification is unaffected there. For any environment where DR contributes significantly to ORLs, the Type~B classification should be revisited; DR-dominated recombination lines would be partially tail-sensitive and could shift toward $T_{\rm eff}$, reducing the measured $R$.

$^{\ddagger}$The Langmuir probe classification depends on the analysis method, but a physical consideration makes standard LP analysis functionally Type~A. The sheath potential ($\sim 3\,k_BT_e$) at the probe surface electrostatically repels the cold electron core; only electrons with sufficient energy to overcome this barrier reach the probe. The sheath thus acts as a threshold gate analogous to the ionization threshold, preferentially sampling the suprathermal tail. Unlike a fixed atomic threshold, however, the sheath potential is self-consistent: a stronger suprathermal tail drives the floating potential more negative, dynamically raising the energy cutoff. This nonlinear feedback makes the LP gate distribution-dependent, but the qualitative effect is preserved; the sheath-selected population always overrepresents the tail relative to the core. A standard exponential fit to the resulting I-V characteristic, interpreted under the Maxwellian assumption, systematically overestimates $T_e$ for any distribution with an enhanced tail. This is the same mechanism that produces the convergence effect in spectroscopy, and it explains why standard LP analysis consistently overestimates $T_e$ relative to Thomson scattering. Full distribution-function fitting to the complete I-V curve can extract $\kappa$ directly (Type~C).

\section{Domain Survey}
\label{sec:domains}

Published data in two domains show the predicted discrepancy pattern between Type~A and Type~B diagnostics. A third domain (planetary nebulae) tests the framework's boundary of applicability.

\subsection{Solar corona}
\label{sec:corona}

The solar corona is the proof of concept, treated in detail by \citet{Edmonds2026}.

\citet{Mercier2015} used the Nan\c{c}ay Radioheliograph to image the quiet Sun at six frequencies between 150 and 450~MHz over 183 quiet days spanning 2004--2011. From the same dataset they extracted two independent products: a brightness temperature $T_b \approx 620$~kK (Type~B: radio bremsstrahlung samples the distribution core) and a hydrostatic scale-height temperature $T_H \approx 1.5$~MK (dependent on ionization and pressure, both tail-dominated). The ratio $R = T_H/T_b = 2.4 \pm 0.3$ is spectrally self-consistent across 150--445~MHz, stable over an eight-year solar cycle, and confirmed by LOFAR at lower frequencies \cite{Vocks2018}.

Turbulent scattering accounts for at most $\sim$20\% of the brightness temperature suppression \cite{Sharma2020}. The remainder implies $\kappa = 2.57^{+0.29}_{-0.19}$ via Eq.~(\ref{eq:inversion}). Five EUV diagnostics (all Type~A) converge on $T_{\rm eff} \approx 1.5$~MK, consistent with the prediction of the convergence principle.

Non-hydrostatic contributions to $R = T_H/T_b$ can be bounded. Solar wind outflow at the quiet Sun base ($v \sim 1$--10~km/s \cite{Cranmer2009}, $M^2 \sim 5 \times 10^{-3}$) contributes $\Delta R < 0.02$. Non-spherical geometry is the dominant systematic: the $\sim$2--3$\times$ quiet-Sun density variation between coronal holes and streamers, mitigated by the azimuthal averaging and 8-year time baseline of \citet{Mercier2015} (rms density dispersions of 7--12\%), gives $\Delta R \lesssim 0.2$ as an upper estimate. Transient activity is negligible after quiet-day selection. Helium abundance uncertainty ($n_{\rm He}/n_{\rm H}$ uncertain by $\pm 50\%$ around the standard coronal value 0.085) contributes $\Delta R \approx \pm 0.18$; higher He pushes $R$ upward (strengthening the non-Maxwellian conclusion). The combined systematic is $\Delta R \lesssim 0.3$ (quadrature of 0.2 geometric, 0.18 helium, 0.02 outflow), small compared to $R - 1 = 1.4 \pm 0.3$. Adding the statistical uncertainty ($\pm 0.3$) in quadrature gives a total uncertainty of $\sim$0.4. In the worst case (all errors conspiring to reduce $R$: $R_{\rm corrected} = 2.4 - 0.4 = 2.0$), $\kappa = 3.0$---still firmly non-Maxwellian.

The Knudsen number in the quiet corona is $\mathrm{Kn} \sim 0.01$--$0.1$ (using the NRL electron collision frequency \cite{Huba2013} and the coronal temperature scale height $L_T \sim 10^8$--$10^9$~cm), above the kinetic threshold. For Active Region cores ($n_e \sim 10^{9}$--$10^{10}$~cm$^{-3}$), $\mathrm{Kn} \lesssim 0.01$, approaching the fluid regime. The generation mechanism is velocity filtration through the transition region potential \cite{Scudder1992a,Scudder1992b}, confirmed by recent kinetic simulations \cite{Barbieri2024,Barbieri2025}.

\textit{Prediction}: Active Region cores should exhibit $R \leq 1.5$ as collisionality restores thermal equilibrium.

Parker Solar Probe provides a direct test. At PSP perihelion (0.17~AU, $\sim$36~$R_\odot$), $n_e \sim 300$~cm$^{-3}$ and $T_{\rm core} \sim 20$~eV give $\lambda_{ee} \approx 0.08$~AU (using $\ln\Lambda \approx 24$ from the NRL low-density expression) and $\mathrm{Kn} \approx 0.1$--$0.8$ (depending on the assumed temperature gradient scale), well above the 0.01 threshold. The framework predicts that non-Maxwellian structure cannot be erased at this collisionality. \citet{Halekas2020} reported ``nearly Maxwellian'' distributions at 0.17~AU, which appears to contradict this prediction. It does not. The ``nearly Maxwellian'' characterization applies to the isotropic core-halo decomposition; the strahl (field-aligned suprathermal beam), which dominates the suprathermal electron fraction near perihelion, is projected out of this decomposition. The strahl is a non-Maxwellian feature that the isotropic analysis does not capture. Subsequent PSP observations from 0.13 to 0.5~AU confirm that the combined suprathermal halo+strahl fraction increases with decreasing heliocentric distance \cite{Abraham2022}. The radial trend in the halo component ($\kappa_{\rm halo} \approx 7.5$ at 0.4~AU decreasing to 3.2 at 3~AU; \citealt{Stverak2009}, as reanalyzed by \citealt{Pierrard2022}), combined with the correlation between low $\kappa$ and low $n_e$ \cite{Pierrard2022}, is the Kn criterion stated observationally: where Coulomb collisions are ineffective (low density, high Kn), phase-space structure survives. The diagnostic evidence for $\kappa \approx 2.5$ ($R = 2.4$) is measured at the EUV formation height ($\sim$1.05--1.3~$R_\odot$); the PSP data confirm rather than challenge the framework at the nearest in situ measurement point.

\subsection{Tokamak scrape-off layer}
\label{sec:sol}

\subsubsection{Knudsen number in the SOL}
\label{sec:sol_kn}

The tokamak scrape-off layer \cite{Pitcher1997,Stangeby2000} has two relevant length scales: the parallel connection length $L_\parallel \sim 20$~m (midplane to divertor target) and the local temperature gradient scale length $L_T = T_e / |\nabla_\parallel T_e|$, which can be as short as 1--5~cm near the target. The fusion community characterizes collisionality via $\nu^* = L_\parallel / \lambda_{ee}$, which is the inverse of $\mathrm{Kn}(L_\parallel)$.

Table~\ref{tab:kn_sol} presents Knudsen numbers computed from representative SOL parameters using the NRL Formulary \cite{Huba2013} electron collision frequency $\nu_{ee} = 2.91 \times 10^{-6}\, n_e \ln\Lambda / T_e^{3/2}$~s$^{-1}$ (with $n_e$ in cm$^{-3}$, $T_e$ in eV). The Coulomb logarithm is evaluated row-by-row from the NRL low-temperature electron--electron expression \cite{Huba2013}, yielding $\ln\Lambda \approx 13.6$ upstream through $\ln\Lambda \approx 7.2$ in the detached divertor. For each case we report both the bulk $\mathrm{Kn}_\parallel = \lambda_{ee}/L_\parallel$ and the tail Kn, where the latter accounts for the $v^4$ scaling of the Coulomb mean free path \cite{Shoub1983}: an electron at velocity $v = \alpha\, v_{\rm th}$ has $\lambda(\alpha) = \alpha^4 \lambda_{\rm bulk}$.

\begin{table}[htbp]
\centering
\caption{Knudsen numbers in the tokamak SOL. $\lambda_{ee} = v_{\rm th}/\nu_{ee}$ with $\nu_{ee}$ from the NRL Formulary \cite{Huba2013} and $v_{\rm th} = \sqrt{2k_BT_e/m_e}$. $\mathrm{Kn}_\parallel = \lambda_{ee}/L_\parallel$ uses the connection length ($L_\parallel = 20$~m). $v/v_{\rm th} = \sqrt{E_{\rm th}/k_BT_e}$ is the velocity of ionizing electrons relative to thermal ($E_{\rm th} = 13.6$~eV for H~I, 47.9~eV for C~III). $\mathrm{Kn}_{\rm tail} = (v/v_{\rm th})^4 \times \mathrm{Kn}_\parallel$.}
\label{tab:kn_sol}
\begin{ruledtabular}
\begin{tabular}{lddddd}
Region & \multicolumn{1}{c}{$n_e$ (cm$^{-3}$)} & \multicolumn{1}{c}{$T_e$ (eV)} & \multicolumn{1}{c}{$\mathrm{Kn}_\parallel$} & \multicolumn{1}{c}{$v/v_{\rm th}$} & \multicolumn{1}{c}{$\mathrm{Kn}_{\rm tail}$} \\
\hline
Upstream & 10^{13} & 100 & 0.75 & 0.69^{\rm a} & 0.17 \\
Mid-SOL & 5 \times 10^{12} & 30 & 0.14 & 0.67^{\rm b} & 0.03 \\
Far SOL & 10^{12} & 15 & 0.18 & 0.95^{\rm b} & 0.15 \\
Divertor (attached) & 5 \times 10^{13} & 10 & 0.002 & 1.2 & 0.004 \\
Divertor (high recycling) & 10^{14} & 3 & 1 \times 10^{-4} & 2.1 & 0.002 \\
Divertor (detached) & 5 \times 10^{13} & 1 & 3 \times 10^{-5} & 3.7 & 0.005 \\
\end{tabular}
\end{ruledtabular}
\smallskip

\noindent$^{\rm a}$C~III threshold (47.9~eV); H~I is sub-thermal at this $T_e$ and uninformative.\\
\noindent$^{\rm b}$H~I threshold (13.6~eV); $E_{\rm th} < k_BT_e$, so ionization is a bulk process and the convergence effect is weak. Divertor rows also use H~I. See text.
\end{table}

First, the upstream and mid-SOL are in the kinetic regime ($\mathrm{Kn}_\parallel = 0.1$--$0.8$) at the \textit{bulk} level: even without the $v^4$ tail enhancement, the electron mean free path is a substantial fraction of $L_\parallel$. The convergence principle applies to bulk transport, not only to tail electrons, in these conditions. Second, for H~I ionization, the ionization threshold lies near or below the thermal energy in the upstream SOL ($T_e \sim 30$--100~eV), so $v/v_{\rm th} < 1$ and $\mathrm{Kn}_{\rm tail} < \mathrm{Kn}_{\rm bulk}$ because the $v^4$ scaling works in reverse for sub-thermal electrons, making them \textit{more} collisional than the bulk average. Ionization in this regime is not tail-dominated and the convergence effect is weak; upstream H~I diagnostics should therefore serve as a natural control group where $R \approx 1$ and standard Maxwellian analysis is safe. Third, in the divertor ($T_e = 1$--10~eV), local collisionality is high ($\mathrm{Kn}_\parallel \ll 0.01$, Table~\ref{tab:kn_sol}). By the local Kn criterion, the divertor plasma should be Maxwellian.

The resolution is \textit{non-local transport}. Suprathermal electrons from the hot upstream SOL stream along field lines toward the divertor target. An upstream electron at $v = 2\,v_{\rm th}$ ($E = 400$~eV in 100~eV plasma) has $\mathrm{Kn}_{\rm tail} = 2^4 \times 0.75 \approx 12$, making it deeply collisionless over $L_\parallel$. These electrons arrive at the divertor with energies far above the local thermal energy, creating non-Maxwellian tails that are not generated locally but are maintained by the upstream source. This is the mechanism identified by kinetic simulations \cite{Batishchev1997,Mijin2021,Power2025}: parallel heat flux is carried by a suprathermal population that outruns local thermalization.

A competing effect acts at the target itself: the sheath potential ($\sim 3\,k_BT_e$) preferentially removes high-energy electrons, establishing a loss cone that depletes the tail. The observed tail enhancement is therefore a balance between the upstream streaming source and the sheath sink, strongest in the mid-SOL and far SOL where streaming dominates, weakest at the target plate.

The framework predicts that the kappa-driven component of the Type~A vs Type~B discrepancy should be largest in the attached divertor and mid-SOL, where streaming electron transport is strongest, and should diminish in deep detachment, where high density collisionalizes even the streaming electrons. Published LP-spectroscopy discrepancies are largest in attached and high-recycling regimes \cite{Lomanowski2015}, but separating the kappa contribution from LP instrumental effects requires independent $\kappa$ measurements in these conditions.

\subsubsection{Published diagnostic discrepancies}
\label{sec:sol_discrepancies}

Multiple devices report systematic disagreement between diagnostics that the convergence framework explains.

\textit{Langmuir probe vs Thomson scattering.}
\citet{Stangeby2000} documented systematic Langmuir probe overestimates of $T_e$ relative to Thomson scattering in the SOL. The most dramatic case is JET-ILW, where \citet{Lomanowski2015} found Langmuir probes overestimated $T_e$ by a factor of 5--10 relative to Balmer and Paschen series spectroscopy in detached plasmas. Both diagnostics have mixed Type classification: LP is sheath-gated (functionally Type~A) but subject to well-documented instrumental artifacts; Balmer spectroscopy in detached divertors has significant recombination contributions (bulk-sensitive, Type~B) but retains tail-sensitive excitation contributions (Type~A). A 5--10$\times$ discrepancy between two partially-Type-A diagnostics cannot be cleanly attributed to a kappa-driven bias, which is bounded at $R < 2$ for any physically plausible $\kappa > 3$. LP overestimation in JET detached divertor conditions is independently documented: \citet{Guillemaut2014} showed that JET-ILW divertor LPs systematically overestimated $T_e$ relative to interpretive EDGE2D-EIRENE modeling, attributing the discrepancy to sheath effects in the detached regime. We do not claim quantitative agreement with the Lomanowski result. \citet{Popov2009} measured bi-Maxwellian electron energy distributions near the last closed flux surface in CASTOR. The bi-Maxwellian structure implies that a Langmuir probe fitting a single Maxwellian to such a distribution would return $T_e$ dominated by the high-energy component, consistent with earlier ASDEX observations where Thomson scattering $T_e$ (bulk-sensitive) was a factor of $\sim$2 below the Langmuir-derived value.

The fusion diagnostic community has identified established instrumental effects that contribute to Langmuir probe overestimates: kinetic sheath expansion, secondary electron emission, and fluctuation-induced rectification \cite{Stangeby2000}. The $\kappa$ framework operates in addition to these: even after correcting for sheath physics and fluctuations, a Maxwellian fit to a non-Maxwellian I-V curve systematically overestimates $T_e$ because the exponential region of a kappa distribution is shallower than Maxwellian.

\textit{Spectroscopic evidence.}
\citet{Linehan2023} found that helium emission spectroscopy gives inaccurately high $T_e$ below 10~eV in TCV detached plasmas, a discrepancy that ``cannot be rectified'' within the standard collisional-radiative model's electron-impact excitation and recombination framework. Other explanations (molecular contributions, opacity effects) may contribute; we note only that the observation is consistent with the convergence framework's prediction. He~I line ratios are same-ion ratios (the ion fraction cancels), but the He~I excitation thresholds ($\sim$21--23~eV) at divertor temperatures ($\sim$2~eV) give $E_{\rm exc}/k_BT \approx 10$, placing excitation deep in the tail-dominated regime (Sec.~\ref{sec:excitation}). If non-Maxwellian tails are present, the line ratios would track $T_{\rm loc}(E_{\rm exc})$ rather than $T_{\rm core}$, producing exactly the observed pattern of systematically high inferred $T_e$.

\textit{Non-Maxwellian distributions measured directly.}
Kinetic simulations by \citet{Batishchev1997} showed strong non-Maxwellian tails forming from non-local parallel transport in the SOL, particularly in detached conditions. More recently, \citet{Mijin2021} confirmed non-Maxwellian distributions in SOL-KiT simulations, and \citet{Power2025} demonstrated that non-local electron transport significantly modifies ionization balance and radiative power loss rates. The role of non-Maxwellian tails in enhancing ionization rate constants was noted explicitly in the divertor physics context by \citet{Krasheninnikov2017}.

These distributions are typically described as ``bi-Maxwellian'' because upstream parallel transport and local recycling produce two distinct electron populations, making a two-temperature decomposition physically motivated. As shown quantitatively in Sec.~\ref{sec:kappa_fit}, a single kappa distribution with $\kappa \approx 2$--10 reproduces the published bi-Maxwellian EEDF shapes to within 3--8\% RMS with one fewer free parameter; the two descriptions are not distinguishable for the diagnostic argument, since both produce the same convergence pattern in Type~A diagnostics.

\textit{Ionization rate anomalies.}
\citet{Power2025} demonstrated that electron distributions with enhanced high-energy tails modify both the ionization balance and the radiative power loss rates in tokamak divertor plasmas, with excitation-driven radiative losses changing by 50--75\%. This is the convergence mechanism discovered independently by the fusion community; ionization and excitation rates track the tail, not the bulk.

\subsubsection{Parametric equivalence of kappa and bi-Maxwellian descriptions}
\label{sec:kappa_fit}

The bi-Maxwellian decompositions cited above use four free parameters $(n_1, T_1, n_2, T_2)$. A natural question is whether a single isotropic kappa distribution, with three free parameters $(n, T_{\rm core}, \kappa)$, can reproduce these shapes. The following exercise is illustrative, not a formal model comparison: we fit kappa distributions to published bi-Maxwellian parameters reconstructed from the literature, not to raw I-V probe data. The exercise tests parametric equivalence (whether a kappa reproduces the shape a bi-Maxwellian describes), not whether it provides a statistically superior fit to the underlying measurements. AIC/BIC model selection does not apply. We fit kappa distributions to the published bi-Maxwellian parameters from four datasets spanning three devices: CASTOR $r = 56$~mm and $r = 66$~mm (EPS 2007 \cite{Popov2007}), COMPASS shot \#2568 at 9~mm from the LCFS, and TJ-II shot \#34531 during NBI heating at 13~mm inside the confined plasma \cite{Popov2015}.

The kappa model reproduces all four bi-Maxwellian EEDFs to within 3--8\% RMS despite having one fewer free parameter, with best-fit values spanning $\kappa \approx 10$ (CASTOR, mildly non-Maxwellian edge) through $\kappa \approx 5$ (COMPASS, near LCFS) to $\kappa \approx 2$ (TJ-II NBI, strongly driven). In the COMPASS case, independent Thomson scattering measurements report $T_e$ consistent with the cold bi-Maxwellian component $T_1 = 6$~eV, a factor of $\sim$2 below the probe-inferred value \cite{Popov2015}, precisely the Type~B vs Type~A split the taxonomy predicts. The fitted kappa core temperature ($T_{\rm core} = 4.6$~eV) is lower still because the kappa tail redistributes energy from the core into the suprathermal wing, giving $T_{\rm core} < T_1 < T_{\rm eff}$. The TJ-II result ($\kappa \approx 2$, marginally above the physical minimum $\kappa > 3/2$) shows where the description strains. Beam injection creates a localized feature in velocity space that no smooth single distribution captures well, consistent with the 8\% RMS residual being the worst of the four fits.

These results are not a substitute for fitting kappa distributions to raw probe I-V curves, which would constitute a proper statistical test. They show something narrower but sufficient: the diagnostic discrepancies reported in the tokamak SOL are quantitatively consistent with kappa-distributed electrons spanning $\kappa \approx 2$--10, and the Thomson--probe temperature split is an independent, published confirmation of the Type~A/Type~B mechanism.

A proper statistical comparison requires fitting both models to raw I-V probe data. We attempted to obtain raw data from COMPASS shot \#2568 (correspondence with IPP Prague, March 2026). The data are not publicly available, and the original experimentalist, Prof.\ Tsviatko Popov, is deceased. We note that the inverse problem (fitting kappa distributions to raw Langmuir probe I-V curves) is well-defined and could be implemented on existing probe analysis codes. The predicted signatures are a shallower-than-Maxwellian exponential region and excess current at high retarding potentials, both of which have been observed empirically but attributed to instrumental effects rather than distribution shape.

\subsubsection{Predicted versus published diagnostic ratios}
\label{sec:predicted_R}

The diagnostic framework makes a testable quantitative prediction: the ratio $R = T_A/T_B$ between any Type~A and Type~B diagnostic applied to the same plasma should equal $T_{\rm eff}/T_{\rm core} = \kappa/(\kappa - 3/2)$ for the local $\kappa$. Table~\ref{tab:predicted_R} compares this prediction against published diagnostic discrepancies using $\kappa$ values derived independently from the EEDF fitting exercise above.

\begin{table}[htbp]
\centering
\caption{Predicted Type~A--to--Type~B diagnostic temperature ratio from EEDF-derived $\kappa$ values (\S\ref{sec:kappa_fit}). $R_{\rm predicted} = \kappa/(\kappa - 3/2)$ assumes $T^* \approx T_{\rm eff}$, valid to $\pm$15--25\% in the SOL regime (Fig.~\ref{fig:error_propagation}). The COMPASS row is the only entry with a published LP--Thomson comparison; the CASTOR rows are untested predictions. Observational uncertainty propagation: at $R = 1.45$, $\delta R \approx \pm 0.3$ gives $\delta\kappa \approx \pm 2.2$ ($\pm$46\%).}
\label{tab:predicted_R}
\begin{ruledtabular}
\begin{tabular}{lcccc}
Device & $\kappa$ & $R_{\rm predicted}$ & Published ratio & Diagnostic pair \\
\hline
COMPASS LCFS & 4.8 & 1.45 & $\sim$2 & LP vs Thomson \\
CASTOR $r=56$~mm & 10 & 1.18 & --- & (prediction) \\
CASTOR $r=66$~mm & 8 & 1.23 & --- & (prediction) \\
\end{tabular}
\end{ruledtabular}
\end{table}

For the COMPASS case (the only dataset with both an independent $\kappa$ measurement and a published LP--Thomson comparison), the framework predicts $R = 1.45$, while the published ratio is approximately 2. The predicted value is smaller, not larger, than the observation. This is the expected direction: LP measurements include well-documented instrumental effects (kinetic sheath expansion, secondary electron emission, fluctuation-induced rectification) that inflate $T_{\rm LP}$ beyond the kappa-driven $T_{\rm eff}$. The residual discrepancy ($R_{\rm predicted} = 1.45$ vs.\ $R_{\rm observed} \approx 2$) is consistent with known LP systematic overestimation from sheath expansion effects, bounded at up to 60\% for standard 4-parameter fits without perimeter correction \cite{Tsui2018}. Correcting published LP data for instrumental effects and checking whether the residual matches the framework's prediction is a direct test. Equivalently, any LP--Thomson discrepancy below $R \approx 1.4$--1.5 in a plasma with measured $\kappa \approx 5$ would falsify the kappa interpretation.

\subsubsection{ITER implications}
\label{sec:iter}

Divertor heat flux models use Spitzer--H\"arm conductivity: $q_{\rm SH} \propto T_e^{7/2} / L_T$. If spectroscopic $T_e$ measures $T_{\rm eff}$ rather than $T_{\rm core}$, the temperature input is biased by a factor $\kappa/(\kappa - 3/2)$. The Spitzer heat flux error scales as this ratio raised to the 7/2 power:

\begin{equation}
\frac{q_{\rm SH}(T_{\rm eff})}{q_{\rm SH}(T_{\rm core})} = \left(\frac{T_{\rm eff}}{T_{\rm core}}\right)^{7/2} = \left(\frac{\kappa}{\kappa - 3/2}\right)^{7/2}
\label{eq:spitzer_error}
\end{equation}

\begin{table}[htbp]
\centering
\caption{Spitzer--H\"arm heat flux bias from using $T_{\rm eff}$ instead of $T_{\rm core}$, computed from raw Spitzer--H\"arm conductivity [Eq.~(\ref{eq:spitzer_error})] without flux limiters or kinetic corrections. See text for discussion of how this bias propagates through flux-limited transport models.}
\label{tab:spitzer}
\begin{ruledtabular}
\begin{tabular}{cccc}
$\kappa$ & $T_{\rm eff}/T_{\rm core}$ & Heat flux bias & Regime \\
\hline
10  & 1.18 & 1.8$\times$ & Mild \\
5   & 1.43 & 3.5$\times$ & Significant \\
3   & 2.0  & 11$\times$  & Order-of-magnitude \\
2.5 & 2.5  & 25$\times$  & Qualitative failure \\
\end{tabular}
\end{ruledtabular}
\end{table}

For $\kappa \leq 2$, the third velocity moment (heat flux) diverges mathematically; fluid transport models are undefined. At $\kappa \approx 3$--5, consistent with the bi-Maxwellian distributions measured in the SOL \cite{Popov2009}, the Spitzer--H\"arm formula overestimates parallel heat flux by a factor of 3--11. The bias also infects flux limiters: substituting $T_{\rm eff}$ for $T_{\rm core}$ inflates the free-streaming ceiling by $(T_{\rm eff}/T_{\rm core})^{3/2}$, a residual 2--4$\times$ even after the limiter caps $q_{\rm SH}$ (Sec.~\ref{sec:heat}). Additionally, \citet{Landi2001} showed that the Spitzer--H\"arm law breaks down at $\kappa \lesssim 5$, and at $\kappa \lesssim 10$ heat flux can reverse direction entirely. The error is not merely quantitative; it can be qualitative.

This is a potential systematic bias in ITER divertor heat load predictions. \citet{Power2023} found up to 50\% heat flux suppression and 98\% enhancement of the sheath heat transmission coefficient in kinetic SOL simulations, qualitatively consistent with the Spitzer--H\"arm breakdown predicted here (though their work frames the effect as non-local transport rather than kappa formation). Determining the actual $\kappa$ in the ITER-relevant SOL parameter space is an empirical question that the diagnostic framework proposed here can resolve.

\subsection{Planetary nebulae}
\label{sec:nebulae}

This section tests the framework's self-consistency by applying the Kn validity criterion to a regime where it predicts inapplicability. Planetary nebulae challenge the boundary of the convergence framework. Unlike the solar corona and tokamak SOL, planetary nebulae are in photoionization equilibrium (PIE), not collisional ionization equilibrium (CIE). The O$^{2+}$ population is created by $> 35$~eV UV photons from the central star; collisional ionization is negligible at $\sim 10^4$~K. The ionization bottleneck of the convergence principle (Sec.~\ref{sec:theorem}) does not apply directly; any kappa-driven diagnostic bias must operate through collisional excitation of forbidden lines, not ionization.

The standard nebular temperature diagnostic, the [O~III] $\lambda 4363/\lambda 5007$ ratio, depends on collisional \textit{excitation} ($\sim$5~eV threshold), not on ionization ($\sim$55~eV); see \citet{StoreySochiBadnell2014} for updated collision strengths, and \citet{StoreySochi2015} for kappa-dependent effective collision strengths. At $T_{\rm core} \approx 0.86$~eV, the excitation threshold gives $E_{\rm exc}/k_BT_{\rm core} \approx 6$, placing excitation in the tail-dominated regime (Sec.~\ref{sec:excitation}). If the electron distribution has suprathermal tails, CELs measure $T_{\rm loc}(E_{\rm exc})$, which exceeds $T_{\rm core}$. Recombination lines (ORLs) remain bulk-sensitive (Type~B) because recombination cross-sections scale as $\sigma \propto E^{-1}$ and sample the thermal peak. The qualitative prediction ($T_{\rm CEL} > T_{\rm ORL}$) is preserved; the quantitative question is whether suprathermal tails survive thermalization at the diagnostic energy.

\subsubsection{Knudsen number in planetary nebulae}
\label{sec:pn_kn}

The bulk electron Knudsen number in planetary nebulae is small: $\mathrm{Kn}_{\rm bulk} \sim 10^{-9}$ to $10^{-8}$ at typical parameters ($n_e \sim 10^2$--$10^4$~cm$^{-3}$, $T_e \sim 10^4$~K, $L \sim 10^{16}$--$10^{17}$~cm). The plasma is firmly fluid. The Shoub $v^4$ mean-free-path scaling improves the picture by orders of magnitude but does not overcome the deficit; both energy thresholds remain below the kinetic boundary.

\begin{figure}[htbp]
\centering
\includegraphics[width=\columnwidth]{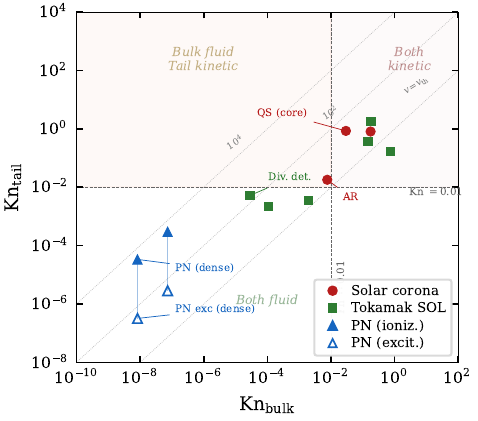}
\caption{Knudsen number map for all three domains. $\mathrm{Kn}_{\rm bulk} = \lambda_{ee}/L$ is the standard electron collisionality; $\mathrm{Kn}_{\rm tail} = (v/v_{\rm th})^4 \times \mathrm{Kn}_{\rm bulk}$ incorporates the Shoub $v^4$ mean-free-path scaling at the relevant threshold energy. Dashed lines mark the $\mathrm{Kn} = 0.01$ kinetic boundary; diagonal dotted lines indicate constant $v^4$ enhancement factors. Plasma parameters for the solar corona follow \citet{Mercier2015}; SOL parameters are from Table~\ref{tab:kn_sol} \cite{Stangeby2000,Pitcher1997}; PN parameters use standard nebular values \cite{Wesson2018}. The three domains occupy physically distinct regions of the map: the solar corona (circles; QS $=$ Quiet Sun, AR $=$ Active Region) is deeply kinetic at both bulk and tail; the tokamak SOL (squares, upstream through detached divertor, ``Div.\ det.'') straddles the boundary, with non-local parallel transport from the hot upstream SOL maintaining suprathermal tails where the bulk is collisional (Sec.~\ref{sec:sol_kn}); planetary nebulae (triangles; filled at the 54.9~eV ionization threshold, open at the 5.35~eV [O~III] excitation threshold) fall deep in the fluid zone at both diagnostic energies, identifying PN as the framework's falsification boundary.}
\label{fig:kn_map}
\end{figure}

O~III \textit{ionization} requires electrons at $E_{\rm th} = 54.9$~eV when $k_BT \approx 0.86$~eV, giving $v/v_{\rm th} \approx 7.99$ and an enhancement factor of $(v/v_{\rm th})^4 \approx 4075$. For a typical PN ($n_e = 10^4$~cm$^{-3}$):
\begin{equation}
\mathrm{Kn}_{\rm tail}^{\rm (ioniz)} = \mathrm{Kn}_{\rm bulk} \times (v/v_{\rm th})^4 \approx 8 \times 10^{-9} \times 4075 \approx 3 \times 10^{-5}
\end{equation}
Even with the $v^4$ enhancement, the tail electrons responsible for O~III ionization are collisional over the full nebular scale ($\mathrm{Kn} \ll 0.01$; Fig.~\ref{fig:kn_map}).

As established above, PNe are in PIE and the diagnostic bias operates through \textit{excitation}, not ionization. The [O~III] $\lambda 4363$ excitation threshold is only 5.35~eV, giving $v/v_{\rm th} = \sqrt{5.35/0.86} \approx 2.5$ and an enhancement factor of $(2.5)^4 \approx 39$:
\begin{equation}
\mathrm{Kn}_{\rm tail}^{\rm (exc)} = \mathrm{Kn}_{\rm bulk} \times 39 \approx 3 \times 10^{-7}
\label{eq:kn_exc}
\end{equation}
The 5~eV electrons that drive the diagnostic bias are even more deeply collisional ($\mathrm{Kn} \sim 10^{-7}$). At this energy, local Coulomb collisions enforce a Maxwellian, consistent with the \citet{DraineKreisch2018} finding that the distribution thermalizes up to $\sim$13--16~eV under spatially homogeneous conditions. The Kn threshold of 0.01 (the standard kinetic-theory boundary above which perturbative corrections to the Maxwellian become order-unity \cite{Braginskii1965,ScudderKarimabadi2013}) is not met at either the diagnostic energy or the ionization energy.

This identifies PNe as the falsification boundary: electrons at all relevant diagnostic energies, from the 5~eV excitation threshold to the 55~eV ionization threshold, are too deeply collisional for macroscopic kappa distributions to survive.

A persistence length calculation confirms this constraint quantitatively. The diffusive persistence length $L_{\rm persist} = v/(\nu\sqrt{3})$, where $\nu$ is the Coulomb collision frequency at speed $v$ with the Shoub $v^{-3}$ scaling, measures how far a non-Maxwellian perturbation at a given energy penetrates before local thermalization erases it. At 5~eV ($v/v_{\rm th} = 2.5$, Shoub $v^4$ enhancement $= 39\times$), $L_{\rm persist} \approx 0.011$~AU at $n_e = 10^2$~cm$^{-3}$ (the most favorable case), shrinking to $1.2 \times 10^{-4}$~AU at $n_e = 10^4$~cm$^{-3}$. The steepest nebular gradient (ionization fronts at cometary knot boundaries \cite{ODell2002}) has a characteristic ionization skin depth $\sim 1/(n_{\rm H}\sigma_{\rm ph}) \sim 10^{13}$~cm ($\sim$0.7~AU) at $n_{\rm H} \sim 10^5$~cm$^{-3}$, a factor of $\sim 60\times$ larger than $L_{\rm persist}$ even at the lowest density. Velocity filtration cannot bridge the gap at the diagnostic energy.

At the ionization threshold (54.9~eV, $v/v_{\rm th} = 8.0$, enhancement $= 4075\times$), $L_{\rm persist} \approx 1.2$~AU in diffuse gas ($n_e = 10^2$~cm$^{-3}$), shrinking to $\sim$0.013~AU at $n_e = 10^4$. Even at the ionization energy, the persistence length is comparable to cometary knot scales only in the most diffuse gas; at typical nebular densities it falls short by a factor of $\sim$5. The uniform collisionality across both thresholds confirms that PNe sit below the kinetic boundary at all diagnostic energies.

\subsubsection{Upper bounds on the kappa contribution}
\label{sec:pn_adf}

Setting $\mathrm{Kn}_{\rm tail}^{\rm (exc)} \gtrsim 0.01$ would require gradient scale lengths $L \lesssim 3 \times 10^{11}$~cm ($\sim$0.02~AU) at $n_e = 10^4$~cm$^{-3}$, far smaller than both the global nebular radius and cometary knot ionization fronts ($\sim$0.7~AU) \cite{ODell2002}. Whether steep gradients at these micro-scales can sustain non-Maxwellian tails at 5~eV against $\sim$30-second local Coulomb thermalization \cite{DraineKreisch2018} is unknown quantitatively. The following are strict upper bounds on the kappa contribution to the abundance discrepancy, contingent on this unestablished micro-scale physics.

If suprathermal tails persist at the excitation energy, the local slope temperature $T_{\rm loc}(E)$ from Eq.~(\ref{eq:Tloc}) gives the CEL-to-ORL temperature ratio. The CEL--ORL discrepancy is $R = T_{\rm loc}/T_{\rm ORL}$, not $T_{\rm eff}/T_{\rm core}$ as in CIE environments; the quantitative extraction of $\kappa$ uses Eq.~(\ref{eq:Tloc}) rather than the inversion formula [Eq.~(\ref{eq:inversion})]. Table~\ref{tab:pne_predict} lists the upper-bound CEL--ORL temperature ratio at the [O~III] $\lambda 4363$ excitation energy ($E_{\rm exc} \approx 5.35$~eV) for a range of $\kappa$ values, together with the approximate abundance discrepancy factor (ADF).

\begin{table}[htbp]
\centering
\caption{Upper bounds on the CEL--ORL temperature ratio and abundance discrepancy factor from the $T_{\rm loc}$ approximation [Eq.~(\ref{eq:Tloc})] at the [O~III] $\lambda 4363$ excitation threshold ($E_{\rm exc} = 5.35$~eV, $T_{\rm core} = 0.86$~eV). These values assume kappa-distributed electrons at the diagnostic energy, a condition not validated by the Knudsen number criterion in bulk nebular gas ($\mathrm{Kn}_{\rm tail}^{\rm (exc)} \approx 3 \times 10^{-7} \ll 0.01$). They represent strict upper bounds contingent on unestablished micro-scale physics (Sec.~\ref{sec:pn_kn}). The ADF is estimated from the emissivity bias $\exp[E_{5007}(T_{\rm core}^{-1} - T_{\rm loc}^{-1})/k_B]$ with $E_{5007} \approx 2.49$~eV. These are saddle-point estimates; full numerical integration of [O~III] excitation rate coefficients with kappa distributions \cite{StoreySochiBadnell2014} may shift individual values by up to $\sim$30\% (cf.\ Table~\ref{tab:rate_check}), though the qualitative trend is robust.}
\label{tab:pne_predict}
\begin{ruledtabular}
\begin{tabular}{cccc}
$\kappa$ & $T_{\rm loc}/T_{\rm core}$ & $T_{\rm loc}$ (K) & ADF \\
\hline
10 & 1.47 & 14\,700 & $\sim$2.5 \\
15 & 1.33 & 13\,300 & $\sim$2.0 \\
20 & 1.25 & 12\,500 & $\sim$1.8 \\
30 & 1.17 & 11\,700 & $\sim$1.5 \\
\end{tabular}
\end{ruledtabular}
\end{table}

For $\kappa \approx 15$, the upper-bound ratio $R = 1.33$ matches the observed CEL--ORL temperature discrepancy in moderate-ADF nebulae (e.g., NGC~7009, NGC~6543), and the upper-bound ADF~$\approx 2$ falls within the observed range for typical objects. This value of $\kappa$ is independently consistent with the \citet{YaoZhang2022} constraint ($\kappa \geq 15$ from UV C~II line ratios) and with the \citet{StoreySochi2013} finding that Maxwellian distributions are not excluded.

The mechanism saturates, and emission weighting tightens the bound further. Even at $\kappa = 3$ (implausibly non-Maxwellian for PN electron densities of $10^2$--$10^4$~cm$^{-3}$), $T_{\rm loc}/T_{\rm core} = 2.3$ and the single-zone ADF reaches only $\sim$5. Moreover, the persistence length analysis above shows that non-Maxwellian tails at 5~eV survive only in diffuse gas ($n_e \lesssim 10^2$~cm$^{-3}$), while [O~III] emission scales as $n_e^2$ and is therefore dominated by denser regions. In a two-zone model where 10--20\% of [O~III] emission originates in diffuse gas with $\kappa \approx 15$, the emission-weighted observable ADF is only $\sim$1.1--1.3, well below the single-zone upper bound of $\sim$2. Extreme abundance discrepancies (ADF~$> 10$, reaching 70--80 in objects like Hf~2-2) are physically unreachable through a kappa distribution alone.

If the micro-scale physics does not sustain tails (the default expectation from the Kn criterion), the kappa contribution is identically zero. This yields a sharpened bifurcation in the ADF population:

\begin{itemize}
\item \textit{Moderate ADF} ($\sim$2--5): the kappa contribution is bounded at ADF~$\approx 1.2$--1.5 after emission weighting, insufficient by itself even under the most favorable micro-scale assumptions. Moderate ADFs require \textit{both} distribution shape effects \textit{and} mild Peimbert-type temperature fluctuations \cite{Peimbert1967}, with neither mechanism alone accounting for the full discrepancy. If kappa tails are absent (as the Kn criterion suggests for bulk gas), Peimbert fluctuations must carry the entire load.
\item \textit{Extreme ADF} ($> 10$): requires genuinely two-component plasma (physically distinct regions at different temperatures and/or compositions), not a single-zone distribution shape effect. The \citet{StoreySochi2014} finding that a two-component Maxwellian strongly outperforms kappa for Hf~2-2 is consistent with this constraint, and the emission-weighting bound makes it robust.
\end{itemize}

\citet{Wesson2018} showed that extreme ADFs correlate with binary central stars, which produce ejected hydrogen-deficient material through common-envelope interactions. The bifurcation maps onto this correlation: extreme ADFs require real inhomogeneity (binarity provides the mechanism), while moderate ADFs may not.

Independent observational constraints are consistent with mild or absent non-Maxwellian tails in PNe. \citet{StoreySochi2013} found Maxwellian distributions best-fit from C~II dielectronic recombination lines, though uncertainties could not exclude kappa. \citet{StoreySochi2014} found a two-component Maxwellian strongly favored over kappa for the extreme-ADF nebula Hf~2-2. \citet{YaoZhang2022} inferred $\kappa \geq 15$ from UV C~II line ratios. These results constrain the PN application to subtle departures ($\kappa \gtrsim 10$--15) if any, independently consistent with the Kn criterion's prediction that 5~eV electrons are thermalized in bulk gas.

The PN application constrains rather than confirms the framework. The diagnostic taxonomy correctly classifies CELs as tail-sensitive and ORLs as bulk-sensitive, but the Kn criterion that validates the framework in the corona and SOL identifies PNe as the domain where the mechanism fails. The kappa contribution to the CEL--ORL discrepancy is bounded above at ADF~$\approx 1.2$--1.5 under maximally favorable micro-scale assumptions. The remaining discrepancy in moderate-ADF objects requires an additional mechanism, most plausibly mild Peimbert-type temperature fluctuations, while extreme ADFs ($> 10$) are definitively beyond the reach of distribution shape effects. The solar and tokamak cases stand on their own physics; the PN domain demonstrates that the framework contains its own falsification criterion.

\citet{Nicholls2012} proposed kappa distributions as an explanation for the abundance discrepancy. The convergence framework identifies the mechanism (threshold-crossing diagnostics sample the tail), but the Kn analysis shows it is not validated in bulk PN gas.

\FloatBarrier

\subsection{Other domains}
\label{sec:other_domains}

Table~\ref{tab:other_domains} summarizes additional environments where the framework applies.

\begin{table}[htbp]
\centering
\caption{Additional plasma environments where the convergence framework applies. Kn$_{\rm tail}$ includes the Shoub $v^4$ scaling for the relevant diagnostic ion.}
\label{tab:other_domains}
\begin{ruledtabular}
\begin{tabular}{lllll}
Domain & $\mathrm{Kn}_{\rm tail}$ & Type~A & Type~B/C & Published $\kappa$ \\
\hline
Solar wind & $\gg 1$ & Charge states & In situ (C) & 2--8 \cite{Stverak2009} \\
Magnetospheres & $0.1$--10 & ENA spectroscopy & In situ (C) & 2--8 \cite{Vasyliunas1968} \\
Io torus & $\sim 0.1$ & UV spectroscopy & Radio (B) & $\sim$2.4 \cite{MeyerVernet1995} \\
Laser plasmas & 0.01--10 & X-ray spectroscopy & Thomson (B) & Unknown \\
AGN jets & $\gg 1$ & X-ray spectroscopy & None & Unknown \\
\end{tabular}
\end{ruledtabular}
\end{table}

The solar wind and planetary magnetospheres provide ground truth. In situ particle detectors (Type~C) measure kappa distributions directly, and charge-state freezing temperatures (Type~A) are known to exceed in situ electron temperatures. The Io plasma torus is a published precedent for velocity filtration generating non-Maxwellian distributions in a non-solar environment \cite{MeyerVernet1995}.

\section{What Breaks the Degeneracy}
\label{sec:break}

\subsection{Thomson scattering}

Thomson scattering measures the one-dimensional electron velocity distribution along the scattering wavevector via Doppler broadening of scattered laser light \cite{Sheffield2011}. For a three-dimensional isotropic kappa distribution [Eq.~(\ref{eq:kappa})], the one-dimensional marginal is \cite{Livadiotis2013}
\begin{equation}
f_{\rm 1D}(v_z) \propto \left(1 + \frac{v_z^2}{\kappa\,\theta^2}\right)^{-\kappa},
\label{eq:TS_1D}
\end{equation}
obtained by integrating Eq.~(\ref{eq:kappa}) over the perpendicular velocity components, which reduces the exponent from $-(\kappa+1)$ to $-\kappa$. The result has power-law wings rather than the exponential cutoff of a Gaussian. Standard Thomson analysis fits a Gaussian to this spectrum. What temperature does the fit report?

The answer depends on which feature of the spectrum dominates the fit. The peak curvature of Eq.~(\ref{eq:TS_1D}) at $v_z = 0$ is $f''(0)/f(0) = -2/\theta^2$, independent of $\kappa$; a Gaussian matched to this curvature reports $T_{\rm core}$ exactly. The variance of Eq.~(\ref{eq:TS_1D}) is $\langle v_z^2\rangle = \kappa\theta^2/(2\kappa - 3)$; a Gaussian matched to this second moment reports $T_{\rm eff}$ exactly. A least-squares Gaussian fit, which captures both the core shape and partial wing information, reports an intermediate value $T_{\rm fit}$ satisfying
\begin{equation}
T_{\rm core} < T_{\rm fit} < T_{\rm eff}\,.
\label{eq:TS_bounds}
\end{equation}
Numerical least-squares fitting of a Gaussian to Eq.~(\ref{eq:TS_1D}) gives $T_{\rm fit}/T_{\rm core} = 1.49$ at $\kappa = 2.5$ (where $T_{\rm eff}/T_{\rm core} = 2.5$), $T_{\rm fit}/T_{\rm core} = 1.38$ at $\kappa = 3$ ($T_{\rm eff}/T_{\rm core} = 2.0$), and $T_{\rm fit}/T_{\rm core} = 1.20$ at $\kappa = 5$ ($T_{\rm eff}/T_{\rm core} = 1.43$). These values are insensitive to the fit window for $v_{\rm max} \geq 3\theta$. In practice, photon noise in the spectral wings forces experimentalists to weight the fit heavily toward the core ($v \lesssim 2\theta$), effectively discarding the power-law tail information entirely. This pushes $T_{\rm fit}$ toward $T_{\rm core}$, making experimental Thomson scattering a purer Type~B diagnostic than the uniform least-squares bound suggests. The observed diagnostic discrepancy $R_{\rm obs} = T_A/T_{\rm fit}$ is therefore a better proxy for $R_{\rm true} = T_{\rm eff}/T_{\rm core}$ than the idealized calculation above implies. The physical origin of the bias is excess kurtosis: the kappa 1D marginal has kurtosis $\beta_2 = 3 + 6/(2\kappa - 5)$ for $\kappa > 5/2$, diverging as $\kappa \to 5/2$.

The consequence for measured diagnostic ratios is quantitative. A published spectroscopic-to-Thomson ratio $R_{\rm obs} = T_A/T_{\rm fit}$ underestimates the true ratio $R = T_{\rm eff}/T_{\rm core}$ because $T_{\rm fit} > T_{\rm core}$. At $\kappa = 3$, $R_{\rm obs} = T_{\rm eff}/T_{\rm fit} = 2.0/1.38 = 1.45$, whereas $R_{\rm true} = 2.0$. Applying the inversion formula [Eq.~(\ref{eq:inversion})] to $R_{\rm obs}$ yields $\kappa_{\rm inferred} \approx 4.8$, substantially overestimating the true $\kappa = 3$. All published Type~A--to--Thomson discrepancies in the tokamak SOL are therefore lower bounds on the true $R$, and $\kappa$ values inferred from them are upper bounds on the true departure from equilibrium. This upper bound is tightest when the Type~A diagnostic operates near the sweet spot (Fig.~\ref{fig:error_propagation}), where $T^* \approx T_{\rm eff}$ to within $\sim$10\%. For high-threshold ions where $T^*$ substantially exceeds $T_{\rm eff}$ (e.g., C~III in the SOL, $T^*/T_{\rm eff} = 1.42$; Table~\ref{tab:rate_check}), the $T^*$ overshoot partially compensates the Thomson undershoot, and the net direction of the combined error depends on the specific ion and $\kappa$ value.

\citet{Milder2019} confirmed this quantitatively in laser-produced plasmas: fitting non-Maxwellian Thomson spectra with a Gaussian assumption produces errors up to 50\% in $T_e$ and 30\% in density. \citet{Milder2021} demonstrated the first complete electron distribution measurement without shape assumptions, using angularly resolved Thomson scattering. Extending this approach to tokamak SOL diagnostics would provide a direct Type~C measurement of the distribution shape and an independent determination of $\kappa$, free from the Gaussian-fit bias quantified above. The principal challenge is signal-to-noise: detecting power-law wings against photon noise requires higher laser energies or longer integration times than standard Gaussian-fit analysis, particularly in the low-density far SOL where the non-Maxwellian effect is expected to be strongest.

Anisotropy strengthens the Type~B classification. The derivation above assumes an isotropic 3D kappa distribution, but in the tokamak SOL the suprathermal tail is carried by electrons streaming along field lines ($T_\parallel \gg T_\perp$). Thomson scattering measures the velocity component along the laser scattering vector \cite{Sheffield2011}, which in standard tokamak geometry is oriented nearly perpendicular to $\mathbf{B}$, making the measurement geometrically blind to the parallel suprathermal tail. Thomson scattering therefore measures $T_{\rm core,\perp}$ almost exactly. The bounds in Eq.~(\ref{eq:TS_bounds}) become more conservative (not less) under anisotropy: if Thomson undersamples the parallel tail, $T_{\rm fit}$ is pushed closer to $T_{\rm core}$, strengthening the conclusion that published $R_{\rm obs}$ values are lower bounds. The perpendicular geometry makes tokamak Thomson scattering a purer Type~B diagnostic than the isotropic analysis suggests.

\subsection{Radio bremsstrahlung}

Optically thick free-free emission measures $T_{\rm core}$ because bremsstrahlung emissivity is dominated by electrons near the thermal velocity, with no ionization threshold involved \cite{Dulk1985,ChiuderiDrago2004,Fleishman2014}. Radio brightness temperature bypasses the ionization gatekeeper entirely. The solar corona case is the exemplar; radio bremsstrahlung is not available as a diagnostic in the tokamak SOL or planetary nebulae.

\subsection{Recombination lines}

Radiative recombination cross-sections scale as $\sigma \propto E^{-1}$ above threshold, so the rate is dominated by slow electrons near the distribution peak. ORL temperatures approximate $T_{\rm core}$ in kappa plasmas. This is why CEL and ORL temperatures disagree in planetary nebulae: CELs track $T_{\rm loc}(E_{\rm exc})$ through tail-dominated excitation (Sec.~\ref{sec:nebulae}); ORLs track $T_{\rm core}$ through recombination.

\subsection{Langmuir probes}

As noted in Sec.~\ref{sec:taxonomy}, the sheath potential at the probe surface acts as a threshold gate, collecting only electrons energetic enough to overcome the $\sim 3\,k_BT_e$ barrier. Standard Maxwellian fit to the resulting exponential region systematically overestimates $T_e$ for kappa distributions because the sheath-selected population has a shallower effective slope than the bulk. Full distribution-function fitting to the complete I-V curve can extract $\kappa$ directly.

The bi-Maxwellian EEDFs measured at CASTOR \cite{Popov2007,Popov2009} and similar devices (Sec.~\ref{sec:sol_discrepancies}) have a clear physical motivation: upstream parallel transport and local recycling produce two distinct electron populations, making a two-temperature fit natural. The kappa-description equivalence of these distributions is demonstrated in Sec.~\ref{sec:kappa_fit}; fitting kappa distributions directly to raw probe I-V curves rather than to published decomposition parameters would provide a proper statistical model comparison and an independent determination of $\kappa$.

\section{Implications}
\label{sec:implications}

\subsection{Heat transport}
\label{sec:heat}

The quantitative consequences of using $T_{\rm eff}$ in place of $T_{\rm core}$ in transport calculations are severe. Spitzer--H\"arm conductivity \cite{SpitzerHarm1953}, $q_{\rm SH} = \kappa_0 T^{7/2}/L_T$, assumes a Maxwellian, collision-dominated plasma \cite{Braginskii1965}. In a kappa-distributed plasma with Kn $> 0.01$, neither assumption holds.

Modern fluid transport codes (SOLPS-ITER \cite{Wiesen2015}, UEDGE \cite{Rognlien1992}) do not apply raw Spitzer--H\"arm conductivity; they employ heat flux limiters of the form $q_\parallel = \min(q_{\rm SH},\, \alpha\, n_e k_B T_e c_s)$ precisely because modelers know SH breaks down in kinetic regimes. The more insidious problem is upstream of the flux limiter: the temperature input itself. If spectroscopic $T_e$ measures $T_{\rm eff}$ rather than $T_{\rm core}$, the upstream boundary conditions and target constraints fed to the transport model are biased before the flux limiter is ever applied. The temperature error alone (Table~\ref{tab:spitzer}) produces biases of 3--25$\times$ in the raw SH formula. Moreover, because the limiter ceiling scales with the ion acoustic speed ($c_s \propto (T_e + T_i)^{1/2}$), a biased $T_e$ inflates the flux cap in addition to inflating $q_{\rm SH}$.\footnote{$T_i$ is often measured independently via CXRS, partially diluting the $T_e$ bias in the flux limiter ceiling but not in $q_{\rm SH}$ itself.} The surviving bias under flux-limited transport can be bounded: the free-streaming limit scales as $n_e k_B T_e c_s \propto T_e^{3/2}$, so substituting $T_{\rm eff}$ for $T_{\rm core}$ inflates the ceiling by $(T_{\rm eff}/T_{\rm core})^{3/2}$. For $\kappa = 2.5$ this is $4.0\times$; for $\kappa = 3$, $2.8\times$; for $\kappa = 5$, $1.7\times$. The raw Spitzer--H\"arm bias (3--25$\times$) is reduced to this residual 2--4$\times$ in the flux-limited regime, but it is not eliminated. The limiter does not correct the bias; it inherits it at reduced amplitude. The biased $T_e$ then propagates further into density reconstructions, ionization source terms, and radiation loss calculations. A subtler consequence is that transport modelers routinely tune empirical radial transport coefficients ($D_\perp$, $\chi_e$, $\chi_i$) to match downstream target data, often from Langmuir probes or spectroscopy. If the target $T_e$ constraint is biased by a kappa tail, modelers will inadvertently distort their inferred radial transport profiles to compensate, embedding the $\kappa$ bias in the empirically fitted cross-field diffusion without any visible diagnostic disagreement. Because the power fall-off length $\lambda_q$ is extracted from these same fitted transport profiles, the bias propagates directly into the parameter that sizes the ITER divertor.

The deeper problem is that Spitzer--H\"arm is derived from a collision-dominated moment closure that breaks down for $\kappa \lesssim 5$ \cite{Landi2001}. At $\kappa \lesssim 10$, heat flux can reverse direction entirely \cite{Landi2001}. At $\kappa \leq 2$, the third velocity moment diverges; the heat flux integral does not converge.

These results apply to both the solar corona and the tokamak SOL. For ITER, the divertor heat load predictions depend on SOL parallel transport modeling. Determining the actual $\kappa$ in the ITER-relevant SOL parameter space, using the diagnostic framework proposed here, is a prerequisite for assessing whether current transport models adequately capture the boundary physics.

\subsection{The diagnostic prescription}
\label{sec:prescription}

For any plasma where $\mathrm{Kn} > 0.01$ (the standard kinetic-theory threshold above which perturbative corrections to the Maxwellian become order-unity \cite{Braginskii1965,ScudderKarimabadi2013}):
\begin{enumerate}
\item Identify which diagnostics in use are Type~A and which are Type~B (Table~\ref{tab:taxonomy}).
\item If $T_A \approx T_B$: either Maxwellian or $\kappa > 10$ ($R < 1.2$, effect negligible).
\item If $T_A > T_B$: compute $R = T_A/T_B$ and extract $\kappa$ via Eq.~(\ref{eq:inversion}). This inversion is a localized approximation valid when the diagnostic threshold sits near the sweet spot of Eq.~(\ref{eq:sweet_spot}); for threshold ratios $E_{\rm th}/k_BT_{\rm core} \gg 5\kappa/(2\kappa - 3)$, full forward modeling with Eq.~(\ref{eq:rate_equivalence}) is required (cf.\ Fig.~\ref{fig:error_propagation} and Table~\ref{tab:rate_check}).
\item If $\kappa < 5$: Spitzer--H\"arm is invalid. Use kinetic transport or flux-limited models.
\item Every diagnostic campaign on a weakly collisional plasma should include at least one Type~B measurement.
\end{enumerate}

\subsection{Cooling functions and energy budgets}

Astrophysical cooling functions assume Maxwellian distributions. If the electron distribution is kappa, radiative loss rates computed from ionization equilibrium at $T_{\rm eff}$ are systematically wrong. This affects coronal energy balance, nebular photoionization models, cluster cooling flow models, and ISM energy budgets. We flag this as a consequence of the framework but do not attempt a quantitative correction, which requires domain-specific forward modeling.

\section{Discussion}
\label{sec:discussion}

\subsection{What this paper does and does not claim}
\label{sec:scope}

This paper claims: (1) ionization-based diagnostics are structurally degenerate in $\kappa$; (2) their convergence on a common temperature is not evidence of Maxwellian equilibrium; (3) the Kn criterion, with the Shoub $v^4$ correction for tail particles, identifies where this matters; (4) the diagnostic taxonomy classifies which methods are affected and which are not; (5) published discrepancies in the corona, tokamak SOL, and planetary nebulae are consistent with the framework.

This paper does not claim: (a) all reported plasma temperatures are wrong (Kn $< 0.01$ plasmas are unaffected); (b) specific $\kappa$ values for specific tokamak experiments (that requires dedicated reanalysis); (c) this invalidates spectroscopic measurements (it reinterprets what they measure: $T_{\rm eff}$, not $T_{\rm core}$); (d) kappa distributions are the only possible non-Maxwellian form (the convergence principle holds for any distribution with a suprathermal tail; kappa is the calculable case).

A methodological assumption throughout is isotropy: the kappa distribution [Eq.~(\ref{eq:kappa})] and all rate calculations assume an isotropic velocity distribution. Real plasmas are often anisotropic: the tokamak SOL has $T_\parallel \gg T_\perp$ due to parallel streaming, and the solar wind strahl is field-aligned. Because collisional ionization depends on electron speed $|v|$, not direction, an anisotropic tail still enhances ionization rates relative to a Maxwellian at the same $T_{\rm core}$, and the qualitative convergence result holds: Type~A diagnostics remain degenerate. However, the mapping from the observed $R = T_A/T_B$ to a single scalar $\kappa$ becomes ambiguous for anisotropic distributions. The inversion formula [Eq.~(\ref{eq:inversion})] should be understood as yielding an effective isotropic-equivalent $\kappa$ that characterizes the diagnostic bias but may not correspond to the $\kappa$ of any single velocity component. Generalizing the framework to bi-kappa or anisotropic distributions is straightforward in principle but requires specifying the diagnostic geometry (e.g., Thomson scattering wavevector orientation relative to $\mathbf{B}$; see Sec.~\ref{sec:break}).

\subsection{Relation to existing work}
\label{sec:relation}

The kappa distribution community \cite{Livadiotis2013,Dudik2015,Dudik2017,Dzifcakova2018,Pierrard2010} has computed line ratios, ionization equilibria, and emission measures under kappa. The fusion diagnostics community has catalogued non-Maxwellian effects in the SOL as ``bi-Maxwellian'' populations and ``non-local transport'' \cite{Batishchev1997,Power2023}; this physics is established, and the present paper does not claim to discover it. The standard collisional-radiative infrastructure for fusion plasma modeling, ADAS \cite{Summers2006}, computes rate coefficients under the Maxwellian assumption; the \citet{HahnSavin2015} decomposition and the KAPPA package \cite{Dzifcakova2015KAPPA} extend these rates to kappa distributions. \citet{Nicholls2012} proposed kappa distributions for nebular abundance discrepancies. The contribution of this paper is not the underlying atomic physics or transport theory, but the structural observation: stating the convergence result as a general principle, introducing the diagnostic taxonomy that classifies which measurements are degenerate in $\kappa$ and which are not, and connecting the independently discovered anomalies across three communities under a single framework.

\citet{DraineKreisch2018} solved the steady-state electron energy distribution in H~II regions and planetary nebulae from first principles, including Coulomb scattering, photoionization, recombination, free-free emission, and collisional excitation. Their central result---that Coulomb thermalization is fast ($\tau_{\rm relax} \approx 30$~s at Orion Nebula densities) and the steady-state distribution is Maxwellian to high accuracy up to $\sim$13~eV (H~II regions) or $\sim$16~eV (PNe), even with continuous injection of 10--20~eV photoelectrons from the radiation field---is a rigorous local calculation and is not in dispute. The persistence length analysis in Sec.~\ref{sec:nebulae} reaches the same conclusion independently at 5~eV: velocity filtration cannot sustain non-Maxwellian tails at the diagnostic energy in bulk PN gas. The D\&K result is spatially homogeneous, however, and does not treat velocity filtration \cite{Scudder1992a} along the steep density gradients present at ionization fronts and cometary knots \cite{ODell2002}. Whether spatial inhomogeneity modifies the picture at micro-scales ($\lesssim 10$~AU) remains open, but the convergence framework does not depend on this question being resolved in its favor: the diagnostic taxonomy applies to whatever tail exists, however small, because threshold-gated and bulk-sensitive diagnostics weight it differently by construction.

Observational tests in PNe (\citealt{StoreySochi2013}; \citealt{StoreySochi2014}; \citealt{YaoZhang2022}; discussed in Sec.~\ref{sec:nebulae}) are consistent with D\&K's prediction. These results constrain the PN application to subtle departures ($\kappa \gtrsim 10$--15) if any, but do not bear on the convergence principle itself, which is structural.

\subsection{Open questions}
\label{sec:open}

\begin{enumerate}
\item What are the actual $\kappa$ values in the ITER-relevant SOL? Can existing Thomson and spectroscopy datasets be reanalyzed with the inversion formula?
\item Does velocity filtration operate at the divertor sheath boundary, analogous to the solar transition region?
\item Can Thomson scattering systems be optimized to detect non-Gaussian wings and extract $\kappa$ directly?
\item Does the $\kappa$ framework resolve the nebular abundance discrepancy quantitatively? The [O~III] excitation subtlety (Sec.~\ref{sec:nebulae}) means the inversion formula cannot be applied directly; a full forward-modeling exercise with kappa ionization equilibria is required.
\item Why is the coronal $\kappa \approx 2$--3 so far from the perturbative predictions of \citet{Cranmer2014} ($\kappa \approx 10$--25)? Velocity filtration is fundamentally non-perturbative; whether the perturbative approach systematically underestimates the departure is unresolved. In situ constraints from PSP confirm non-Maxwellian structure at 0.17~AU (Sec.~\ref{sec:corona}), consistent with the Kn criterion, but do not resolve the quantitative discrepancy with perturbative theory.
\item How does velocity-space anisotropy ($T_\parallel \neq T_\perp$) affect the inversion formula? In the tokamak SOL, suprathermal tails are primarily parallel; in the solar wind, the strahl is field-aligned. The isotropic $\kappa$ inferred from $R$ is an effective parameter whose relation to component-resolved $\kappa_\parallel$ and $\kappa_\perp$ depends on the diagnostic geometry.
\end{enumerate}

\section{Conclusions}
\label{sec:conclusions}

\begin{enumerate}
\item Ionization-based diagnostic convergence is structurally circular for any plasma with Kn~$> 0.01$. $N$ diagnostics downstream of collisional ionization report $T_{\rm eff}$; their agreement is a single measurement repeated, not $N$ independent confirmations.

\item The diagnostic taxonomy (Type~A/B/C) classifies which methods are degenerate in $\kappa$ and which break the degeneracy. The taxonomy is applicable across solar, fusion, and astrophysical plasma physics.

\item The convergence framework explains published diagnostic discrepancies in the solar corona ($R = 2.4$, $\kappa \approx 2.5$ \cite{Mercier2015}) and the tokamak SOL (Langmuir--Thomson discrepancies of factors 2--10, non-Maxwellian distributions measured directly). In planetary nebulae, the framework predicts the qualitative CEL--ORL pattern but the Knudsen number and persistence length at both the diagnostic energy (5~eV) and the ionization energy (55~eV) do not validate velocity filtration as the sustaining mechanism; electrons at all relevant energies are firmly collisional over nebular scales. The kappa contribution to moderate ADFs is bounded above at $\sim$1.2--1.5 after emission weighting, contingent on unestablished micro-scale physics at ionization fronts and cometary knots. If micro-scale tails are absent (the default expectation from the Kn criterion), Peimbert-type temperature fluctuations must account for the full moderate-ADF discrepancy. Extreme ADFs ($> 10$) are definitively beyond the reach of distribution shape effects alone.

\item For plasmas with $\kappa \approx 3$--5, the raw Spitzer--H\"arm formula with spectroscopic $T_e$ as input overestimates parallel heat flux by factors of 3--25$\times$. Flux-limited transport models inherit the biased temperature through their boundary conditions. This is directly relevant to ITER divertor heat load predictions.

\item Every diagnostic campaign on a weakly collisional plasma should include at least one Type~B measurement. The ratio of Type~A to Type~B temperatures directly yields $\kappa$ via $\kappa = 3R/[2(R-1)]$.
\end{enumerate}

\section*{Acknowledgements}

The author acknowledges the open-science community for making the literature, data, and computational tools accessible enough for independent contribution. The author thanks the Institute of Plasma Physics, Czech Academy of Sciences (IPP Prague) for correspondence regarding COMPASS probe data availability.

Computational verification of analytical results (Knudsen number calculations, thermalization timescales, ADF predictions) and preparation of script-generated figures were assisted by Claude (Anthropic). All physical arguments, derivations, and scientific conclusions are the author's own.

\section*{Data Availability}

This paper interprets previously published observational results and performs Knudsen number calculations from published plasma parameters. No new experimental data were generated. Raw probe I--V characteristics from COMPASS tokamak shot \#2568 were requested from IPP Prague (Czech Academy of Sciences); the data are not publicly archived and could not be obtained. The Knudsen number calculations are available from the author on request.


\end{document}